\documentstyle[12pt]{article}
\title{ General Gauge Field Theory  And Its Application }
\author{Wu Ning \\[2mm]
Division 1, Institute of High Energy Physics\\
Beijing 100039, P.R.China }
\date{ }

\begin{document}
\maketitle
\begin{abstract}
A gauge field model, which simultaneously has strict local gauge
symmetry and contains massive general gauge bosons, is discussed in this paper. The
model has $SU(N)$ gauge symmetry. In order to introduce the mass term of
gauge fields directly without violating the gauge symmetry of the theory, two
sets of gauge fields will be introduced into the theory. After some
transformations, one set of gauge fields obtain masses and another set of
gauge fields keep massless. In the limit $\alpha \longrightarrow 0$ or
$\alpha \longrightarrow \infty$, the gauge field model discussed in this
paper will return to Yang-Mills gauge field model. Finally, some
applications of this model are discussed.

\end{abstract}


\section{ Introduction}
~~~~ Yang and Mills founded 
non-Abel gauge field theory in 1954\lbrack 1 \rbrack. Since then, 
gauge field theory has been extensively applied to elementary particle 
theories. Now,  it is generally believed that four kinds of fundamental 
interactions, i.e. strong interactions, electromagnetic interactions, 
weak interactions and gravitation, are all gauge interactions.  
From theoretical point of view, the 
requirement of  gauge invariant determines the forms of interactions. 
But for Yang-Mills gauge theory, if lagrangian has strict local 
gauge symmetry, the masses of  gauge fields must be zero. 
On the other hand,  physicists 
found that the masses of intermediate bosons are very large 
in the forties\lbrack 2 \rbrack. After introducing spontaneously symmetry 
breaking and Higgs mechanism,
Glashow\lbrack 3 \rbrack, Weinberg\lbrack 4 \rbrack 
and Salam\lbrack 5 \rbrack  founded the well-known unified electroweak 
standard model. The standard model is consonant 
well with experiments and  intermediate bosons $W^{\pm}$ and $Z^0$
have already been found by experiments. But
Higgs particle has not been found by
experiments until now. We know that Higgs particle is 
necessitated by the standard model. Whether Higgs particle exists in nature? If there were
no Higgs particle, how should we construct the unified electroweak model?
\\

A possible way to solve this problem is to use general gauge field 
theory\lbrack 6 \rbrack.
The general gauge field theory not only has strict local gauge symmetry, 
but also contains massive general gauge bosons. 
Using general gauge field theory, we
could construct an electroweak model which contains no Higgs 
particle\lbrack 7 \rbrack. 
\\

\section{ The lagrangian of the model}

~~~~ Suppose that $N$ 
fermion fields $\psi _l(x)~(l=1,2, \ldots ,N)$ form a 
multiplet of matter fields. The state of matter fields is 
denote as:  \\     
$$
\psi (x) =\left ( 
\begin{array}{c}
\psi_1 (x) \\
\psi_2 (x) \\
\vdots \\
\psi_N (x)
\end{array}
\right ) 
\eqno(2.1)
$$
The system has $SU(N)$ symmetry. The representative matrices
of generators of $SU(N)$ group are denoted by $T_i ~(i=1,2, 
\ldots, N^2-1)$. They satisfy:
$$
\lbrack T_i ~ ,~ T_j \rbrack = i f_{ijk} T_k
\eqno(2.2)
$$
$$
Tr( T_i  T_j ) = \delta_{ij} K.
\eqno(2.3)
$$
A general element of the $SU(N)$ group can be expressed as:
$$
U=e^{-i \alpha ^i T_i}
\eqno(2.4)
$$
with $\alpha ^i$ the real group parameters. U is a unitary $N \times N$
matrix.
\\

We need two kinds of gauge fields $A_{\mu}(x)$ and
$B_{\mu}(x)$. They can be expressed as linear 
combinations of generators :
$$
A_{\mu}(x) = A_{\mu} ^i (x) T_i
\eqno(2.5a)
$$
$$
B_{\mu}(x) = B_{\mu} ^i (x) T_i.
\eqno(2.5b)
$$
where $A_{\mu}^i (x)$ and $B_{\mu}^i (x)$ are component fields.
In the present model, there are two gauge fields corresponds 
to one gauge symmetry.
Corresponds to two kinds of gauge fields,
there are two kinds of  gauge covariant derivatives
$$
D_{\mu} = \partial _{\mu} - ig A_{\mu}
\eqno(2.6a)
$$
$$
D_{b \mu} = \partial _{\mu} + i \alpha g B_{\mu}.
\eqno(2.6b)
$$
The strengths of gauge fields are:
$$
\begin{array}{ccl}
A_{\mu \nu} & = & \frac{1}{-i g} \lbrack D_{\mu} ~,~ D_{\nu} \rbrack
\\
& = & \partial _{\mu} A_{\nu} - \partial _{\nu} A_{\mu}
- i g \lbrack A_{\mu} ~,~ A_{\nu} \rbrack
\end{array}
\eqno(2.7a)
$$
$$
\begin{array}{ccl}
B_{\mu \nu} &=& \frac{1}{i \alpha g} \lbrack D_{b \mu} ~,~ D_{b \nu}
\rbrack
\\
& = & \partial _{\mu} B_{\nu} - \partial _{\nu} B_{\mu}
+ i \alpha g \lbrack B_{\mu} ~,~ B_{\nu} \rbrack. 
\end{array}
\eqno(2.7b)
$$
Similarly, $A_{\mu \nu}$ and $B_{\mu \nu}$ can also be
expressed as linear combinations of generators:
$$
A_{\mu \nu} = A_{\mu \nu}^i T_i
\eqno(2.8a)
$$
$$
B_{\mu \nu}= B_{\mu \nu}^i T_i. 
\eqno(2.8b)
$$
where
$$
A_{\mu \nu}^i = \partial _{\mu} A_{\nu}^i - \partial _{\nu} A_{\mu}^i
+g f^{ijk} A_{\mu}^j    A_{\nu}^k
\eqno(2.9a)
$$
$$
B_{\mu \nu}^i = \partial _{\mu} B_{\nu}^i - \partial _{\nu} B_{\mu}^i
- \alpha g f^{ijk} B_{\mu}^j    B_{\nu}^k .
\eqno(2.9b)
$$
\\

 The lagrangian density of the model is:
$$
\begin{array}{ccl}
\cal L &= &- \overline{\psi}(\gamma ^{\mu} D_{\mu} +m) \psi \\
&&-\frac{1}{4K} Tr( A^{\mu \nu} A_{\mu \nu} )   
-\frac{1}{4K} Tr( B^{\mu \nu} B_{\mu \nu} ) \\
&&-\frac{\mu ^2}{2K ( 1+ \alpha ^2)} 
Tr \left \lbrack (A^{\mu}+\alpha B^{\mu})( A_{\mu}+\alpha B_{\mu} )
\right \rbrack
\end{array}  
\eqno(2.10)
$$
where $\alpha$ is a constant. 
\\

Local gauge transformations are
$$
\psi \longrightarrow \psi ' = U \psi ,
\eqno(2.11)
$$
$$
A_{\mu} \longrightarrow U A_{\mu} U^{\dag}
-\frac{1}{ig}U \partial _{\mu}U^{\dag}
\eqno(2.12)
$$
$$
B_{\mu} \longrightarrow U B_{\mu} U^{\dag}
+\frac{1}{i \alpha g}U \partial _{\mu}U^{\dag}
\eqno(2.13)
$$
then,
$$
A_{\mu \nu} \longrightarrow U A_{\mu \nu} U^{\dag}
\eqno(2.14)
$$
$$
B_{\mu \nu} \longrightarrow U B_{\mu \nu} U^{\dag}
\eqno(2.15)
$$
$$
( A_{\mu} + \alpha B_{\mu} ) \longrightarrow
 U ( A_{\mu} + \alpha B_{\mu} ) U^{\dag}
\eqno(2.16)
$$
It is easy to prove that the lagrangian is invariant under the above local
gauge transformations. Therefore the model has strict
local gauge symmetry. In the above lagrangian, we 
have introduced mass terms of gauge fields without
violating the gauge symmetry of the model.

\section{ Masses of general gauge fields}

~~~~ If we select $A_{\mu}$ and $B_{\mu}$ as basis, the mass matrix of gauge
fields is:
$$
M  = \frac{1}{1+\alpha ^2} \left (
\begin{array}{cc}
\mu ^2  &  \alpha \mu^2  \\
\alpha \mu ^2  &  \alpha^2 \mu^2
\end{array}
\right ) 
\eqno(3.1)
$$
The masses of gauge fields are given by
eigenvalues of mass matrix. Mass matrix has two eigenvalues:
$$
m^2_1 = \mu ^2 ~~,~~  m^2_2 = 0.
\eqno(3.2)
$$
The wave functions of physical particles are given by
eigenvectors of mass matrix. Two eigenvectors of mass matrix are:
$$
\left (
\begin{array}{c}
{\rm cos } \theta  \\  {\rm sin} \theta
\end{array}
\right)
~ , ~~
\left (
\begin{array}{c}
- {\rm sin } \theta  \\  {\rm cos} \theta
\end{array}
\right) 
\eqno(3.3)
$$
where
$$
{\rm  cos} \theta =  \frac{1}{\sqrt{1+\alpha ^2}}
  ~~,~~
{\rm  sin} \theta =  \frac{\alpha}{\sqrt{1+\alpha ^2}} .
\eqno(3.4)
$$
The wave functions of physical particles are defined as:
$$
C_{\mu}={\rm cos}\theta A_{\mu}+{\rm sin}\theta B_{\mu}
\eqno(3.5a)
$$
$$
F_{\mu}=-{\rm sin}\theta A_{\mu}+{\rm cos}\theta B_{\mu}.
\eqno(3.5b)
$$
The above transformations  are pure field
transformations. They can be regarded as 
redefinition of gauge fields. They do not affect
the symmetry of the lagrangian. So, the local gauge symmetry of the lagrangian
is still strictly preserved
after the above transformations. \\

After the above transformations, 
the lagrangian density changes into
$$
{\cal L} = {\cal L}_0 + {\cal L}_I ,
\eqno(3.6)
$$
$$
{\cal L}_0= - \overline{\psi}(\gamma ^{\mu} \partial _{\mu} +m) \psi
-\frac{1}{4} C^{i \mu \nu}_0 C^i_{0 \mu \nu}   
-\frac{1}{4} F^{i \mu \nu}_0 F^i_{0 \mu \nu}
-\frac{\mu ^2}{2} C^{i \mu} C^i_{\mu}.
\eqno(3.7)
$$
$$
\begin{array}{ccl}
{\cal L}_I & = & i g \overline{\psi} \gamma ^{\mu} ( {\rm cos}\theta C_{\mu}
 - {\rm sin}\theta F_{\mu} )  \psi  \\
&&
- \frac{{\rm cos}2 \theta}{2 {\rm cos} \theta} g f^{ijk}C_0^{i \mu \nu}
C^j_{\mu} C^k_{\nu}
+\frac{{\rm sin} \theta}{2 } g f^{ijk}F_0^{i \mu \nu} F^j_{\mu} F^k_{\nu}
\\
&&
+\frac{{\rm sin} \theta}{2 } g f^{ijk}F_0^{i \mu \nu} C^j_{\mu} C^k_{\nu}
+ g {\rm sin} \theta f^{ijk} C_0^{i \mu \nu} C^j_{\mu} F^k_{\nu}  \\
&&
- \frac{1 - \frac{3}{4}{\rm sin}^2 2 \theta}{4 {\rm cos}^2 \theta} g^2
f^{ijk} f^{ilm}  C^j_{\mu} C^k_{\nu} C^{l \mu} C^{m \nu}  \\
&&
- \frac{{\rm sin}^2  \theta}{4} g^2 f^{ijk} f^{ilm} F^j_{\mu} F^k_{\nu} F^{l
\mu} F^{m \nu}  \\
&&
+ g^2 {\rm tg} \theta {\rm cos} 2 \theta f^{ijk} f^{ilm} C^j_{\mu} C^k_{\nu}
C^{l \mu} F^{m \nu}
\\
&&
- \frac{{\rm sin}^2  \theta}{2} g^2 f^{ijk} f^{ilm} ( C^j_{\mu} C^k_{\nu}
F^{l \mu} F^{m \nu}   \\
&&
~~~~~~~~~~~
+ C^j_{\mu} F^k_{\nu} F^{l \mu} C^{m \nu}+ C^j_{\mu} F^k_{\nu} C^{l \mu}
F^{m \nu}) .
\end{array}
\eqno(3.8)
$$
From free lagrangian ${\cal L}_0$, we could see  that the mass
of gauge field $C_{\mu}$ is $\mu$ and the mass of gauge field $F_{\mu}$ is
zero.
$$
m_c = \mu ~~,~~ m_F = 0 .
\eqno(3.9)
$$
\\

Local gauge transformations of $C_{\mu}$ and $F_{\mu}$
respectively are:
$$
C_{\mu} \longrightarrow U C_{\mu} U^{\dag}
\eqno(3.10)
$$
$$
F_{\mu} \longrightarrow U F_{\mu} U^{\dag}
+\frac{1}{i g {\rm sin}\theta } U \partial _{\mu}U^{\dag}
\eqno(3.11)
$$
The form of local gauge transformation of $C_{\mu}$ is homogeneous. 
But obviously, $C_{\mu}$ is not an ordinary vector field or an ordinary
matter field, because $C_{\mu}$ is a linear combination
of the  standard gauge fields
$A_{\mu}$ and $B_{\mu}$ and transmits gauge interactions between matter
fields. So, we call $C_{\mu}$ general gauge field for the moment. \\

\section{ Yang-Mills Limits}

~~~~ In a proper limit, general gauge field theory can return to 
Yang-Mills gauge field theory. There are two kinds of Yang-Mills limits. \\

The first kind of Yang-Mills limit corresponds to very small parameter
$\alpha$. Let:
$$
\alpha \longrightarrow 0 ,
\eqno(4.1)
$$
then
$$
{\rm cos} \theta = 1 ~~,~~ {\rm sin}\theta =0.
\eqno(4.2)
$$
$$
C_{\mu} = A_{\mu} ~~,~~ F_{\mu}=B_{\mu} .
\eqno(4.3)
$$
In this case, the lagrangian density  becomes
$$
\begin{array}{ccl}
{\cal L} & = & - \overline{\psi} \lbrack \gamma ^{\mu} ( \partial _{\mu}
- i g  C^i_{\mu} T^i ) +m \rbrack \psi \\
&&-\frac{1}{4}  C^{i \mu \nu} C^i_{\mu \nu}
-\frac{1}{4}  F^{i \mu \nu} F^i_{\mu \nu} - \frac{\mu ^2}{2} C^{i \mu}
C^i_{\mu} .
\end{array}
\eqno(4.4)
$$
We could see that, only massive gauge field directly interacts with matter field.
This limit corresponds to the case that gauge interactions are mainly
transmitted by massive gauge field.
But if $\alpha$ strictly vanishes, the lagrangian does not
have gauge symmetry and the theory is not renormalizable.
\\

The second kind of Yang-Mills limit corresponds to very large parameter
$\alpha$. Let
$$
\alpha \longrightarrow \infty ,
\eqno(4.5)
$$
then
$$
{\rm cos} \theta = 0 ~~,~~ {\rm sin}\theta =1.
\eqno(4.6)
$$
$$
C_{\mu} = B_{\mu} ~~,~~ F_{\mu}= - A_{\mu} .
\eqno(4.7)
$$
Then, the lagrangian density  becomes
$$
\begin{array}{ccl}
{\cal L} & = & - \overline{\psi} \lbrack \gamma ^{\mu} ( \partial _{\mu}
+ i g  F^i_{\mu} T^i ) +m \rbrack \psi \\
&&-\frac{1}{4}  F^{i \mu \nu} F^i_{\mu \nu}
-\frac{1}{4}  C^{i \mu \nu} C^i_{\mu \nu} - \frac{\mu ^2}{2} C^{i \mu}
C^i_{\mu} .
\end{array}
\eqno(4.8)
$$
In this case, only massless gauge field  directly interacts with matter
fields. This limit corresponds to the case that gauge interactions 
are mainly transmitted by massless gauge field.   \\

In the particles' interaction model, 
the parameter $\alpha$ should be finite,
$$
0 < \alpha < \infty.
\eqno(4.9)
$$
In this case, both massive gauge field and massless gauge field directly
interact with matter fields, and gauge interactions are transmitted 
by both of them.    \\

\section{Is the theory renormalizable}

~~~~ We know that the propagator of a massive vector field usually has 
the following form:
$$
\Delta_{F \mu \nu} =
\frac{-i}{k^2 + \mu^2 - i \varepsilon}
\left ( g_{\mu \nu} + \frac{k_{\mu} k_{\nu}}{\mu^2} \right ).
\eqno(5.1)
$$
So, when we let
$$
k \longrightarrow \infty
\eqno(5.2)
$$
then,
$$
\Delta_{F \mu \nu}
\longrightarrow const .
\eqno(5.3)
$$
According to the power counting law, a massive vector field model 
is not renormalizable in most case.
Though general gauge field theory
contains massive vector fields, it is renormalizable. The key reason is that
the lagrangian has local gauge symmetry\lbrack 8 \rbrack.  \\

The general gauge field theory has
maximum local $SU(N)$ gauge symmetry. When we quantize the general gauge
field theory in the path integral formulation, we must select gauge
conditions first\lbrack 9 \rbrack.  In order to make gauge transformation
degree of freedom completely fixed, we must select two gauge conditions
simultaneously: one is for massive gauge field $C_{\mu}$ and another is for
massless gauge field $F_{\mu}$.
For example, if we select temporal gauge
condition for massless gauge field $F_{\mu}$:
$$
F_4 = 0 ,
\eqno(5.4)
$$
there still exists remainder gauge transformation degree of freedom, because
temporal gauge condition is unchanged under the following local gauge
transformation:
$$
F_{\mu} \longrightarrow U F_{\mu} U^{\dag}
+\frac{1}{i g {\rm sin}\theta } U \partial _{\mu} U^{\dag}
\eqno(5.5)
$$
where
$$
\partial_t U = 0~, ~~~~~ U = U(\stackrel{\rightarrow}{x}) .
\eqno(5.6)
$$
In order to make this remainder gauge transformation degree of freedom 
completely fixed, we had better select another gauge condition for gauge
field $C_{\mu}$. For example, we could select Lorentz gauge condition
for gauge field $C_{\mu}$:
$$
\partial^{\mu} C_{\mu} = 0.
\eqno(5.7)
$$
\\

If we select two gauge conditions simultaneously,
when we quantize the
theory in path integral formulation, there will be two gauge fixing terms in
the effective lagrangian. The effective lagrangian could be written as:
$$
{\cal L}_{eff} = {\cal L}
- \frac{1}{2 \alpha_1} f_1^a f_1^a
- \frac{1}{2 \alpha_2} f_2^a f_2^a
+ \overline{\eta}_1 M_{f1} \eta_1
+ \overline{\eta}_2 M_{f2} \eta_2
\eqno(5.8)
$$
where
$$
f_1^a = f_1^a (F_{\mu}) ~,~~~~~
f_2^a = f_2^a (C_{\mu})
\eqno(5.9)
$$
We could select 
$$
f_2^a = \partial^{\mu} C_{\mu}^a .
\eqno(5.10)
$$
In this case, the propagator of massive gauge field $C_{\mu}$ is 
$$
\Delta_{F \mu \nu}^{ab} (k) =
\frac{-i \delta^{ab}}{k^2 + \mu^2 - i \varepsilon}
\left (g_{\mu \nu} -(1- \frac{1}{\alpha_2})
\frac{k_{\mu} k_{\nu}}{k^2 - \mu^2/ \alpha_2} \right).
\eqno(5.11)
$$
If we let $k$ approach infinity, then
$$
\Delta_{F \mu \nu}^{ab} (k)
\sim \frac{1}{k^2} .
\eqno(5.12)
$$
According to the power counting law,
general gauge field theory is a kind of renormalizable theory.
\\

The local $SU(N)$ gauge symmetry will also give a Ward-Takahashi identity
which will eventually make the theory renormalizable.
So, in the renormalization of the general gauge field theory, 
local gauge symmetry
plays the following two important roles: 1) to make the propagators of 
massive gauge bosons have the renormalizable form; 2) to give a
Ward-Takahashi identity which plays a key role in the renormalization of the
general gauge field theory.  \\

\section{Electroweak model without Higgs particle}

~~~~ As an example, let's discuss electroweak interactions of leptons.
For the sake of convenience, let $e$ 
represent leptons $e,\mu$ or $\tau$, and $\nu$ represent the corresponding 
neutrinos $\nu _e, \nu_{\mu}$ or $\nu_{\tau}$. According to the standard 
model,  $e$ and $\nu$ form left-hand doublet $\psi_L$ which has 
$SU(2)_L$ symmetry and right-hand singlet $e_R$. Neutrinos have no 
right-hand singlets. The definitions of these states are:
$$
\psi_L =\left ( 
\begin{array}{c}
\nu  \\
e
\end{array}
\right )_L
~~,~~~ y= -1
\eqno(6.1)
$$
$$
e_R
~~~~~,~~~ y= -2 ,
\eqno(6.2)
$$
$y$ is the quantum number of weak hypercharge Y. 
\\

Four gauge fields are needed in the new electroweak
theory. They are two non-Abel gauge fields $F_{1 \mu}$ 
and $F_{2 \mu}$ corresponding to the $SU(2)_L$ symmetry and two Abel 
gauge fields $B_{1 
\mu}$ and $B_{2 \mu}$ corresponding to the $U(1)_Y$ 
symmetry.
The strengths of four gauge fields are respectively defined as:
$$
F_{1 \mu \nu} = \partial _{\mu} F_{1 \nu} - \partial _{\nu} F_{1 \mu} 
- i g \lbrack F_{1 \mu} ~~,~~    F_{1 \nu} \rbrack ,
\eqno(6.3a)
$$
$$
F_{2 \mu \nu} = \partial _{\mu} F_{2 \nu} - \partial _{\nu} F_{2 \mu} 
+ i g {\rm tg} \alpha \lbrack F_{2 \mu} ~~,~~    F_{2 \nu} \rbrack ,
\eqno(6.3b)
$$
$$
B_{m \mu \nu} = \partial _{\mu} B_{m \nu} - \partial _{\nu} B_{m \mu} 
 ~~~,~~~~(m=1,2) .
\eqno(6.3c)
$$
\\

In order to introduce symmetry breaking and masses
of all fields, a vacuum potential is needed. 
It has mass dimension. It has no kinematic energy term in the 
lagrangian. So, it has no dynamical degree of freedom. The coupling between
vacuum potential and matter fields can be regarded as a kind of 
interactions between vacuum and matter fields.
\\

The lagrangian density of the model is :
$$
{\cal L} = {\cal L} _l + {\cal L} _g + {\cal L} _{v-l} ,
\eqno(6.4)
$$
where
$$
{\cal L }_l= - \overline{\psi}_L \gamma ^{\mu} 
(\partial _{\mu}+ \frac{i}{2}g \prime B_{1 \mu} -ig F_{1 \mu} ) \psi _L
- \overline{e}_R \gamma ^{\mu} 
(\partial _{\mu}+ ig \prime B_{1 \mu} ) e_R 
\eqno(6.5)
$$
$$
\begin{array}{l}
{\cal L}_g = -\frac{1}{4}  F^{i \mu \nu}_1 F^i_{1 \mu \nu} 
- \frac{1}{4} F^{i \mu \nu}_2 F^i_{2 \mu \nu} 
-\frac{1}{4}  B^{\mu \nu}_1 B_{1 \mu \nu}  
-\frac{1}{4}  B^{\mu \nu}_2 B_{2 \mu \nu}  \\
~~ - v^{\dag} 
\left \lbrack  {\rm cos} \theta _W ( {\rm cos} \alpha F_1^{\mu}+{\rm 
sin}\alpha F_2^{\mu}) -
{\rm sin}\theta _W ( {\rm cos}\alpha B_1^{\mu}+{\rm sin}\alpha 
B_2^{\mu} ) \right \rbrack \\
~~~~\cdot 
\left \lbrack  {\rm cos} \theta _W ( {\rm cos} \alpha F_{1 \mu}+{\rm 
sin}\alpha F_{2 \mu}) -{\rm 
sin}\theta _W ( {\rm cos}\alpha B_{1 \mu}+{\rm sin}\alpha B_{2 \mu} ) 
\right \rbrack
v
\end{array}
\eqno(6.6)
$$
$$
{\cal L} _{v-l} = -f (\overline{e}_R v^{\dag} \psi _L +\overline{\psi}_L v 
e_R) ,
\eqno(6.7)
$$
where $f$ is a dimensionless parameter , $\alpha$ is a constant, $g, ~g'$
are coupling constants and $\theta_W$ are Weinberg angle. $v$ is the vacuum
potential. 
\\

In the original lagrangian density, $v$
has gauge transformation degree of 
freedom. But in our real physical world, the state of 
vacuum can not be varied freely and the  properties of vacuum are rather 
stable, it has no gauge transformation degree of freedom. In  the local 
inertial coordinate system,  vacuum is invariant under space-time  
translation. So $v$ is well-distributed, it is a constant.
Suppose that $v$ has 
the following value
$$
v =\left ( 
\begin{array}{c}
{\rm v}_1\\
{\rm v}_2
\end{array}
\right ),
\eqno(6.8)
$$
where ${\rm v}_1$ and ${\rm v}_2$ satisfy the following relation:
$$
{\rm v}_1^2 + {\rm v}_2^2 = \mu^2 /2 ,
\eqno(6.9)
$$
Make a global $SU(2)_L$ gauge transformation 
so as to make $v$ 
change into the following form
$$
v =\left ( 
\begin{array}{c}
0\\
\mu / \sqrt{2}
\end{array}
\right ).
\eqno(6.10)
$$
When $v$ takes fixed value, the symmetry of the lagrangian is broken
simultaneously. 
\\

Gauge fields $F_{1 \mu} , ~ F_{2 \mu},~ 
B_{1 \mu}$ and $B_{2 
\mu}$ are not eigenvectors of mass matrix.
In order to obtain eigenvectors of 
mass matrix,  we will  make the following two sets of 
transformations of fields. The first set of transformations are:
$$
W_{\mu}={\rm cos}\alpha F_{1 \mu}+{\rm sin}\alpha F_{2 \mu}
\eqno(6.11a)
$$
$$
W_{2 \mu}=-{\rm sin}\alpha F_{1 \mu}+{\rm cos}\alpha F_{2 \mu}
\eqno(6.11b)
$$
$$
C_{1 \mu}={\rm cos}\alpha B_{1 \mu}+{\rm sin}\alpha B_{2 \mu}
\eqno(6.11c)
$$
$$
C_{2 \mu}=-{\rm sin}\alpha B_{1 \mu}+{\rm cos}\alpha B_{2 \mu} .
\eqno(6.11d)
$$
The second set of transformations are:
$$
Z_{\mu}= {\rm sin}\theta _W C_{1 \mu}-{\rm cos}\theta _W W^3_{ \mu}
\eqno(6.12a)
$$
$$
A_{\mu}= {\rm cos}\theta _W C_{1 \mu}+{\rm sin}\theta _W W^3_{ \mu}
\eqno(6.12b)
$$
$$
Z_{2 \mu}= {\rm sin}\theta _W C_{2 \mu}-{\rm cos}\theta _W W^3_{2 
\mu}
\eqno(6.12c)
$$
$$
A_{2 \mu}= {\rm cos}\theta _W C_{2 \mu}+{\rm sin}\theta _W W^3_{2 
\mu}. 
\eqno(6.12d)
$$
\\

After all these transformations, the lagrangian densities
of the model change into:
$$
\begin{array}{ccl}
{\cal L}_l +{\cal L}_{v-l} &= & - \overline{e} (\gamma ^{\mu} 
\partial _{\mu}+ \frac{1}{\sqrt{2}} f \mu ) e 
-\overline{\nu}_L \gamma ^{\mu} \partial _{\mu}\nu _L\\
&&
+\frac{1}{2} \sqrt{g^2 + {g \prime}^2} {\rm sin}2\theta_W  
j^{em}_{\mu} 
 ( {\rm cos}\alpha A^{\mu}- {\rm sin}\alpha  A^{\mu}_2  )  \\
&&
- \sqrt{g^2 + {g \prime}^2} j^{z}_{\mu} 
( {\rm cos}\alpha Z^{\mu} - {\rm sin}\alpha Z_2^{ \mu} ) \\
&&
+ \frac{\sqrt{2}}{2} ig \overline{\nu}_L \gamma ^{\mu} e_L 
( {\rm cos}\alpha W_{\mu}^{+} - {\rm sin}\alpha W_{2 \mu}^{+} )  \\
&&
+ \frac{\sqrt{2}}{2} ig \overline{e}_L \gamma ^{\mu} {\nu}_L 
( {\rm cos}\alpha W_{\mu}^{-} - {\rm sin}\alpha W_{2 \mu}^{-} )
\end{array}
\eqno(6.13)
$$
$$
\begin{array}{ccl}
{\cal L}_g &= &-\frac{1}{2}  W^{+ \mu \nu}_0  W^{-}_{0 \mu \nu} 
-\frac{1}{4}  Z^{\mu \nu} Z_{ \mu \nu} 
-\frac{1}{4}  A^{\mu \nu} A_{ \mu \nu} 
 \\
&&-\frac{1}{2}  W^{+ \mu \nu}_{2 0}  W^{-}_{2 0 \mu \nu} 
-\frac{1}{4}  Z^{\mu \nu}_2  Z_{2  \mu \nu} 
-\frac{1}{4}  A^{\mu \nu}_2  A_{2  \mu \nu} 
\\
&& -\frac{\mu ^2}{2}  Z^{\mu }  Z_{ \mu } 
-\mu ^2 {\rm cos}^2 \theta _W  W^{+ \mu}  W^{-}_{\mu} +{\cal L}_{g 
I}
\end{array} ,
\eqno(6.14)
$$
where ${\cal L}_{g I}$ only contains interaction terms of gauge fields.
Defnitions of field strengths are:
$$
W^{\pm}_{m \mu } = \frac{1}{\sqrt{2}} (W^1_{m \mu} \mp i W^2_{m 
\mu} )
~~( m=1,2, ~W_{1 \mu} \equiv W_{\mu}  ) .
\eqno(6.15)
$$
$$
W^{\pm}_{m0 \mu \nu} = \partial _{\mu} W^{\pm}_{m \nu} 
- \partial _{\nu} W^{\pm}_{m \mu} 
~~( m=1,2, ~W^{\pm}_{1 \mu} \equiv W^{\pm}_{\mu}  ) ,
\eqno(6.16)
$$
$$
Z_{m \mu \nu} = \partial _{\mu} Z_{m \nu} 
- \partial _{\nu} Z_{m \mu} 
~~( m=1,2, ~Z_{1 \mu} \equiv Z_{\mu}  ) ,
\eqno(6.17)
$$
$$
A_{m \mu \nu} = \partial _{\mu} A_{m \nu} 
- \partial _{\nu} A_{m \mu} 
~~( m=1,2, ~A_{1 \mu} \equiv A_{\mu}  ) .
\eqno(6.18)
$$
The currents in the above lagrangian are defined as:
$$
j_{ \mu }^{em} = -i \overline{e} \gamma_{\mu} e
\eqno(6.19)
$$
$$
j_{ \mu }^{Z} =j_{\mu}^{3} - {\rm sin}^2 \theta_W  j_{\mu}^{em}
= i \overline{\psi}_L  \gamma_{\mu} \frac{\tau ^3}{2} \psi_L
- {\rm sin}^2 \theta_W  j_{\mu}^{em} .
\eqno(6.20)
$$
\\

From the above lagrangian, we could see that the mass of fermion $e$ 
is $\frac{1}{\sqrt{2}} f 
\mu$, the mass of neutrino is zero, the masses of charged intermediate gauge
bosons $W^{\pm}$ are 
$\mu {\rm cos} \theta_W$, the mass of neutral intermediate gauge boson $Z$ 
is $\mu = 
\frac{m_W}{{\rm cos} \theta_W}$ and all other gauge fields are massless. 
That is
$$
m_e =  \frac{1}{\sqrt{2}} f \mu~~~,~~~
m_{\nu}=0
\eqno(6.21)
$$
$$
m_W =  \mu {\rm cos} \theta_W ~~~,~~~
m_Z = \mu = \frac{m_W}{{\rm cos} \theta_W}
\eqno(6.22)
$$
$$
m_A= m_{A2}=m_{W2}=m_{Z2}=0
\eqno(6.23)
$$
It is easy to see that, in this model, the expressions of the masses of
fermions 
and intermediate gauge 
bosons are the same as those in the  standard model. \\

In a proper limit, the present model will approximately return to the
standard model. 
Suppose that  parameter $\alpha$ is much smaller than 1,
$$
\alpha \ll 1 ,
\eqno(6.24)
$$
then, in the leading term approximation,
$$
{\rm cos}\alpha \approx 1~~~,~~~
{\rm sin}\alpha \approx 0 .
\eqno(6.25)
$$
In this case, the lagrangian density for fermions becomes:
$$
\begin{array}{ccl}
{\cal L}_l +{\cal L}_{v-l} &= & - \overline{e} (\gamma ^{\mu} 
\partial _{\mu}+ \frac{1}{\sqrt{2}} f \mu ) e 
-\overline{\nu}_L \gamma ^{\mu} \partial _{\mu}\nu _L\\
&&
+{\rm e}  j^{em}_{\mu} A^{\mu}- \sqrt{g^2 + {g \prime}^2} 
j^{z}_{\mu} 
Z^{\mu}\\
&&
+ \frac{\sqrt{2}}{2} ig \overline{\nu}_L \gamma ^{\mu} e_L 
W_{\mu}^{+}
+ \frac{\sqrt{2}}{2} ig \overline{e}_L \gamma ^{\mu} {\nu}_L 
W_{\mu}^{-}
\end{array}
\eqno(6.26)
$$
where 
$$
{\rm e} = \frac{g  g'}{\sqrt{g^2 + {g' }^2}} 
\eqno(6.27)
$$
We see ${\cal L}_l +{\cal L}_{v-l}$ is the same as the corresponding 
lagrangian density in the standard model.
In this approximation, except 
for the terms concern Higgs particle, the 
lagrangian of the model discussed in this paper is almost the same as that
of the standard model: they 
have the same mass relation of  intermediate gauge bosons, the same 
charged currents and neutral 
current, the same electromagnetic current, the same coupling constant of
electromagnetic 
interactions, the same effective coupling constant of weak interactions
$\cdots$ etc.. On the other 
hand, we must see that there are two fundamental differences between the 
new electroweak model 
and the standard model: 1) there is no Higgs particle in the new electroweak
model, so there are no 
interaction terms between Higgs particle and leptons, quarks or 
gauge bosons; 2)compare with the standard 
model, we have introduced two sets of gauge fields in the new electroweak 
model. These new gauge bosons are all massless.
\\

From the above lagrangian, we could see that 
there is no Higgs particle
exist in the present electroweak theory. 
Except for Higgs particle and interactions between 
Higgs particle and leptons, the new
electroweak model keeps almost all other dynamical properties of the
standard model. 
Because the theoretical predictions of the standard model coincide well with
experimental results, we could anticipate that the parameter $\alpha$ will be
very small.
Though vacuum potential $v$ is very like Higgs field, 
they have essential differences. The most important difference is that 
Higgs field has kinematical energy terms but vacuum 
potential has no kinematical energy terms.
In the standard model, masses of all fields,
include quark fields, lepton fields and gauge fields, are generated from
their interactions with Higgs field. In the present model, we could think
that masses of all fields are generated from their interactions with vacuum.
If the  parameter $\alpha$ is small, the cross section
caused by the interchange of massless gauge bosons will be extremely small.
So, there will exists no contradictions between high energy experiments 
and the new electroweak model.
\\

\section{ Comments}

~~~~ In general gauge field theory, if the local gauge symmetry of the
lagrangian is strictly preserved, the mass of general gauge field can be
non-zero. The interaction properties between matter fields and gauge fields 
of the general gauge field theory are the same as those of Yang-Mills gauge
field theory.
In a proper limit, the general gauge field theory can return to
Yang-Mills gauge field theory. It keeps Yang-Mills gauge field theory as
one of its special case.   \\

Because the lagrangian has strict local gauge symmetry, general gauge
field theory is renormalizable\lbrack 8 \rbrack.   \\

If we apply general gauge field theory 
 to electroweak interactions, we could construct
an electroweak model in which Higgs mechanism is avoided. So, Higgs particle
is not a necessary part of an acceptable electroweak model\lbrack 7 \rbrack.

\section*{Reference:}
\begin{description}
\item[\lbrack 1 \rbrack]  C.N.Yang, R.L.Mills, Phys Rev {\bf 96} (1954) 191
\item[\lbrack 2 \rbrack]  T.D.Lee, M Rosenbluth, C.N.Yang, Phys. Rev. {\bf
75} (1949) 9905
\item[\lbrack 3 \rbrack]  S.Glashow, Nucl Phys {\bf 22}(1961) 579
\item[\lbrack 4 \rbrack]  S.Weinberg, Phys Rev Lett {\bf 19} (1967) 1264
\item[\lbrack 5 \rbrack]  A.Salam, in Elementary Particle Theory,
eds.N.Svartholm(Almquist and Forlag, Stockholm,1968)
\item[\lbrack 6 \rbrack]  Ning Wu, General gauge field theory
(hep-ph/9802236 and hep-ph/9805453 )
\item[\lbrack 7 \rbrack]  Ning Wu, A new model for electroweak interactions
(hep-ph/9802237 and hep-ph/9806454)
\item[\lbrack 8 \rbrack]  Ning Wu, The Renormalization of the general
Gauge Field Theory  (in preparation)
\item[\lbrack 9 \rbrack]  Ning Wu, Quantization of the general gauge field
theory,   (in preparation)
\end{description}

\end{document}